\documentclass[conference]{IEEEtran}
\IEEEoverridecommandlockouts
\usepackage{cite}
\usepackage{amsmath,amssymb,amsfonts}
\usepackage{textcomp}
\usepackage{multirow, makecell}
\usepackage{pifont}
\usepackage{subfig}
\usepackage{graphicx}
\usepackage{colortbl}
\usepackage{balance}
\usepackage{xcolor}
\usepackage[space]{grffile}
\usepackage{lscape}
\usepackage{caption}
\usepackage{commath}
\usepackage{tablefootnote}
\usepackage{tikz}
\usepackage{algorithm}
\usepackage{algpseudocode}
\usepackage{xcolor}
\usepackage{listings}
\usepackage{enumitem}
\usepackage{url}

\usepackage{geometry}
\def\BibTeX{{\rm B\kern-.05em{\sc i\kern-.025em b}\kern-.08em
    T\kern-.1667em\lower.7ex\hbox{E}\kern-.125emX}}

\newcommand{\ourmethod}{\textit{PoHAR}}

\begin{document}

\title{\ourmethod{}: Understanding Hyperlocal Human Activities with Pollution Sensor Networks}
\author{\IEEEauthorblockN{Prasenjit Karmakar, Karthik Reddy, Sandip Chakraborty}
\IEEEauthorblockA{\textit{Department of Computer Science and Engineering, Indian Institute of Technology Kharagpur, India.} \\
\{prasenjitkarmakar52282, vkr2471\}@gmail.com, sandipc@cse.iitkgp.ac.in
}
}

\maketitle

\begin{abstract}
Low-cost air quality sensors are becoming ubiquitous in our daily lives as public awareness of air pollution continues to grow, and people take measures to monitor and improve the air they breathe indoors. Besides the standard operation of these sensors, fluctuations in environmental parameters can be leveraged to understand human behavior and activities in indoor spaces. Unlike traditional audio-visual, Radio Frequency, and inertial sensors, air quality sensors are easily scalable to a household, are privacy-preserving, and more economical. Such distributed sensor networks must jointly make decisions to monitor indoor occupants for downstream smart home and healthcare applications. However, due to low processing power, memory, and energy, they often struggle to maintain distributed data consensus and identify activity-affected sensor groups for accurate on-device inference. In this paper, we propose \ourmethod{} framework that implements: (i) a conflict-free replicated data primitive for data sharing, (ii) a hierarchical clustering for ESP32 to detect activity-affected sensor groups with a self-supervised distance metric, and (iii) a leader-based group inference with off-the-shelf ML classifiers, enabling the sensor network to collaboratively detect hyperlocal indoor activities. Our extensive experiments demonstrated on-device activity detection, achieving 97.41\% accuracy for indoor activity and 99.68\% for cooking activity, using off-the-shelf ML models with latency below 34 microseconds.
\end{abstract}

\begin{IEEEkeywords}
    Sensor Networks, Embedded ML, HAR.
\end{IEEEkeywords}

\section{introduction}
In modern indoor environments, accurately detecting and responding to multiple concurrent human activities requires sensor networks that are both intelligent and resource-efficient~\cite{robertson2006general,stork2012audio,fang2016airsense}. While traditional sensing modalities such as cameras~\cite{ribeiro2005human,lin2008human}, microphones~\cite{ntalampiras2018transfer,cristina2024audio}, Radio Frequency (RF) signals~\cite{sen2024continuous,xu2025evaluating}, and inertial wearables~\cite{jain2022collossl,abedin2021attend} have been widely explored for Human Activity Recognition (HAR), their deployment in real-world homes remains limited. Cameras and microphones raise significant privacy concerns, RF requires specialized hardware, and wearables are not scalable. In contrast, low-cost air quality sensors~\cite{fmi2023market,karmakar2024indoor} have become ubiquitous in households, driven by growing public awareness of indoor pollution and the increasing availability of consumer-grade air monitors. Beyond their primary purpose of pollutant tracking, fluctuations in air pollutants encode rich environmental signatures that reflect human activities~\cite{karmakar2024exploring,karmakar2024exploiting} such as cooking, cleaning, ventilation, or movement across rooms, making them a promising, privacy-preserving modality for indoor HAR.

However, harnessing the full potential of such a system introduces unique challenges. Indoor pollutants often form localized hotspots and propagate non-uniformly depending on room structure, ventilation, and the nature of activities~\cite{karmakar2024exploring,karmakar2024indoor}. Consequently, sensors placed in different rooms or even different corners of the same room may capture distinct activity-induced signals, making spatial diversity an inherent characteristic of air-quality-based sensing. Prior work on multi-sensor and multi-view HAR has explored combining data from heterogeneous sensors to achieve richer scene understanding~\cite{gokarn2025ra,xie2019pix2vox,jain2022collossl}. However, such aggregation is largely centralized, leading to high communication overhead when sharing high-fidelity data such as video, audio, or RF channel measurements with a central processor. Moreover, these methods do not account for localized environmental impact, where only a subset of sensors is affected by a given activity, a common characteristic in indoor air quality sensing. This leads to reduced accuracy and an inability to detect multiple simultaneous activities across different indoor zones.

To address spatially varying influence across sensor nodes, researchers in wireless sensor networks have studied distributed clustering and group formation techniques such as HEED~\cite{younis2004heed}, DWEHC~\cite{ding2005distributed}, and DEEC~\cite{qing2006design}. While these energy-aware and similarity-driven approaches demonstrate the feasibility of decentralized grouping, they struggle to scale in dense deployments because communication overhead scales quadratically with the number of nodes~\cite{taherkordi2008communication}. Moreover, these methods rely on static or hand-crafted distance metrics that fail to capture the latent structure of pollution-induced variations. Recent works~\cite{karmakar2024exploring,karmakar2024exploiting} show that pollutant spread creates dynamic, activity-dependent sensor clusters, but existing systems lack mechanisms to detect these affected sensor groups at runtime.

In this paper, we proposed \ourmethod{} that implements a self-supervised (SSL)~\cite{zhang2022self} similarity-aware partition-based hierarchical clustering algorithm~\cite{wang2022pack} for ESP32 microcontrollers to detect affected sensor groups using a distributed conflict-free replicated set-data (set-CvRDT) primitive and RAFT-based~\cite{ongaro2014raft} leader election and group inference for hyperlocal indoor activity recognition. \ourmethod{} is tested with an in-house air quality sensor network deployed in household settings. These sensor devices capture pollutants such as CO\textsubscript{2}, VOCs, and particulate matter, along with humidity and temperature. We conducted comprehensive experiments to evaluate \ourmethod{} framework. We investigate SSL embedding quality using t-SNE. Further, we evaluated Set-CvRDT through multiple scenarios, including concurrent operations and frequent updates, achieving consistent state convergence with minimal convergence latency (i.e., 90\,\textmu s for 5-node consensus), evaluated the on-device clustering (i.e, 13 iterations to reduce 50 nodes to 3 clusters), and leader election with various node configurations and failure scenarios, demonstrating reliability. Finally, we assessed machine learning (ML) models on the leader ESP32, achieving over 97.41\% accuracy for household and 97.28\% for cooking activity recognition. Our key contributions are:
\begin{enumerate}

    \item \textbf{Conflict-free Replicated Set: }We designed a distributed set data structure (Set-CvRDT) for consistent data sharing in a distributed sensor network under node failure. 
    
    \item \textbf{Pollution-aware Clustering of Sensor Nodes: }We implemented a hierarchical clustering algorithm~\cite {wang2022pack} for ESP32 microcontrollers that groups sensors based on SSL embeddings~\cite{zhang2022self} extracted from pollution measurements. This helps improve activity prediction by removing bias from unrelated sensors.
    
    \item \textbf{On-device Hyperlocal Prediction: }We deploy machine learning models directly on ESP32 leader devices for hyperlocal activity recognition. We achieve over 97.41\% accuracy for indoor activity and 97.28\% for cooking activity recognition, with a latency of 34 $\mu$s.
\end{enumerate}
\section{Related Work}
We review the related work in three areas: (i) data modalities, (ii) approaches in human activity recognition (HAR) in contrast to air quality data, and (iii) decision making in distributed sensor networks. Details are as follows.

\subsection{Data Modalities in Human Activity Recognition}
In the HAR literature, researchers primarily used privacy-invasive visual data~\cite{ribeiro2005human,robertson2006general,lin2008human} and acoustic signals~\cite{stork2012audio,ntalampiras2018transfer} to identify human activities. Such systems have limited application indoors, such as in smart homes and healthcare settings, due to privacy concerns. Recently, we have seen the use of non-invasive modalities, such as Radio Frequency (RF)~\cite{singh2019radhar,zhao2023cubelearn,zeng2022multi,sen2024continuous,xu2025evaluating}, that demonstrate HAR capabilities using point-cloud data. Moreover, inertial wearables~\cite{jain2022collossl,abedin2021attend} are also explored in this context. However, the primary drawback of such approaches is that they lack understanding of human behavior beyond movements unless paired with privacy-invasive visual data. For instance, without the visual feed, such systems can only detect a person stir-frying a food item, disregarding the exact item being cooked. Therefore, we propose using air pollution data~\cite{fang2016airsense,karmakar2024exploring,karmakar2024indoor} to capture environmental changes induced by indoor activities.

\subsection{Approaches in Human Activity Recognition}
With video and audio data, researchers focus on learning-based feature extractors~\cite{cristina2024audio,yan2026skelformer,bukht2025review,stork2012audio} to design classifiers. Recent advancements have shifted towards spatio-temporal approaches~\cite{rusu20113d,choy20194d}, along with PointNet~\cite{qi2017pointnet,qi2017pointnet++} and Point-GNN~\cite{shi2020point}, to efficiently process RF-based non-invasive channel state, Doppler, and point-cloud data. Such approaches are very power hungry and require labeled data. We propose using a low-power self-supervised learning (SSL)~\cite{chen2020simple,zhang2022self} framework that leverages readily available air quality data with sparse HAR labels~\cite{karmakar2024indoor}. 

Moreover, with competitive modalities, limitations persist regarding the RF's angular resolution, the camera's FoV, and the microphone's distance. Towards this, researchers have aggregated viewpoints~\cite{gokarn2025ra,xie2019pix2vox,jain2022collossl} from multiple sensors to enrich the model's input data. However, such aggregation leads to significant network load when sharing high-fps audio-visual or RF data with a central server. We propose sharing intermediate low-dimensional SSL-based embeddings from each air quality sensor to a leader node to improve network efficiency.

\subsection{Distributed Decision Making}
Recent studies~\cite{karmakar2024exploring,karmakar2024exploiting} have shown that air pollution forms local hotspots indoors, which are ventilated, trapped, or spread to adjacent rooms depending on human activities like turning on a fan, opening doors or windows, etc. As a result, proximate air quality sensors~\cite{karmakar2024indoor} are affected by the pollution-generating activity. With the spread of such pollutants, a dynamic sensor cluster can pick up environmental changes. This becomes more evident in the case of simultaneous indoor activities. However, identifying such dynamic clusters at run-time is taxing. HEED~\cite{younis2004heed} introduced a hybrid approach that combines a node's residual energy with similarity metrics. DWEHC \cite{ding2005distributed} extended this with multilevel clusters and weighted metrics, reducing intra-cluster energy by 40\% compared to HEED. DEEC~\cite{qing2006design} addressed heterogeneous networks using adaptive probabilities based on residual-to-average energy ratios. However, these methods face scalability issues~\cite{robertson2006general}, as communication overhead grows quadratically with the number of nodes. In contrast, hierarchical methods~\cite{gilpin2013efficient,wang2022pack} scale efficiently. We implemented a partition-based distributed agglomerative hierarchical clustering algorithm~\cite{wang2022pack} for ESP32 microcontrollers on air quality sensors, performing distance-based partitioning and distance-aware merging of intermediate SSL embeddings to dynamically identify the affected sensor cluster. Lastly, each sensor cluster performs a leader-based~\cite{ongaro2014raft} HAR inference with off-the-shelf ML classifiers.
\section{methodology}
\begin{figure*}
    \centering
    \includegraphics[width=0.95\linewidth]{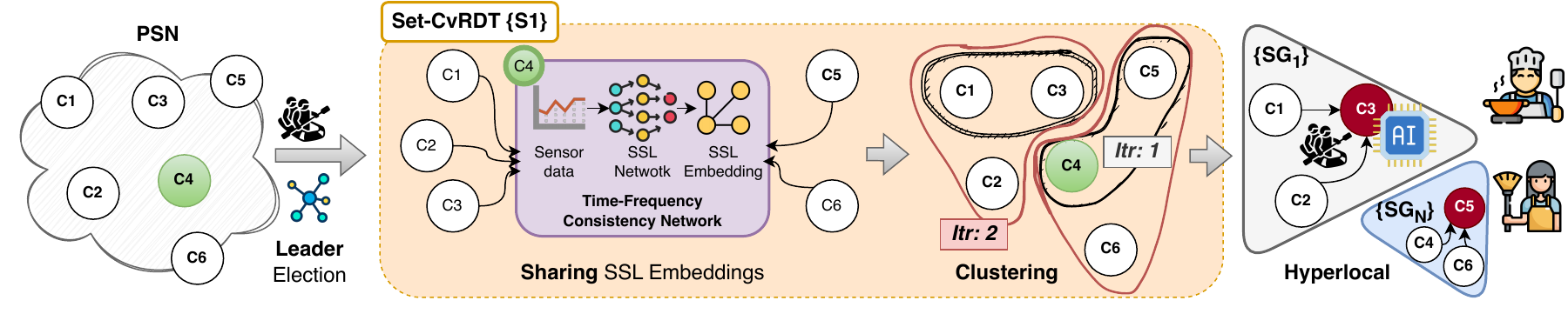}
    \caption{System overview of \ourmethod{}.}
    \label{fig:system}
\end{figure*}

We illustrate the overall workflow of the \ourmethod{} framework in \figurename~\ref{fig:system}. The air quality sensors collaborate to identify activity-affected sensor groups and perform hyperlocal activity recognition. As shown in the diagram, a Pollution Sensor Network (PSN) with multiple nodes (C1–C6) selects the leader (i.e., C4) using the (i) \textit{RAFT election mechanism}~\cite{ongaro2014raft}. Next, each sensor extracts a (ii) \textit{self-supervised learning (SSL) embedding} from its local air quality time-series data using a neural network pretrained with time–frequency consistency~\cite{zhang2022self} on unlabelled air quality data. These low-dimensional embeddings are shared across the network using a (iii) \textit{Set-CvRDT primitive}, ensuring consistent, conflict-free exchange without requiring centralized synchronization. The leader executes the (iv) \textit{hierarchical clustering} algorithm, which iteratively partitions and merges the SSL embeddings based on distances to identify activity-affected sensor groups. Finally, the identified sensor groups, each corresponding to a different indoor activity, perform their own leader-based (v) \textit{hyperlocal inference} using off-the-shelf ML models. For example, one sensor group captures cooking, while another captures a cleaning activity. Thus, \ourmethod{} ensures fine-grained HAR by allowing only the affected sensors to participate in activity inference. Details are as follows.

\subsection{Distributed Leader Election}
The \ourmethod{} framework relies on a lightweight, robust leader election mechanism to coordinate leader-based clustering and inference across resource-constrained sensor nodes. We implemented the RAFT consensus algorithm~\cite{ongaro2014raft}, which provides a fault-tolerant method for electing a stable leader in a distributed setting. RAFT is well-suited for ESP32 microcontrollers due to its simplicity, deterministic behavior, and low communication overhead. Each pollution sensor maintains a local state as \textit{follower}, \textit{candidate}, or \textit{leader} and participates in periodic heartbeat and timeout-driven elections. A sensor transitions to the candidate state when it does not receive a heartbeat from an existing leader within a randomized timeout window. Candidates then request votes from neighboring sensors, and the sensor that obtains a majority becomes the next leader. RAFT ensures that only a single leader is active at any given time, even in the presence of message loss or intermittent connectivity. The elected leader updates the Set-CvRDT and broadcasts it across the sensor network to share local SSL embeddings for pollution-aware clustering.

\subsection{SSL-based Embeddings}
We implemented a time–frequency consistency (TF-C) based SSL model~\cite{zhang2022self} to generate intermediate embeddings from raw air pollution time-series data at each sensor. Indoor activities such as cooking and sleeping often lead to rapid spikes, gradual rises, or periodic fluctuations. To capture these complementary cues without requiring labeled activity data, each sensor processes its local multivariate air quality stream using a lightweight neural encoder (see \figurename~\ref{fig:system}) trained with an SSL objective~\cite{zhang2022self}: augmentations of the same signal segment (e.g., jittering, masking, or frequency perturbations) must produce embeddings that remain close in the latent space, while embeddings from unrelated segments should diverge. By enforcing agreement between the time-domain and frequency-domain views of the same pollution window, the model learns robust low-dimensional representations that encode semantically meaningful information, suitable for downstream clustering and activity inference under noisy sensor readings and diverse indoor conditions.

\begin{figure}
    \centering
    \includegraphics[width=1.0\linewidth]{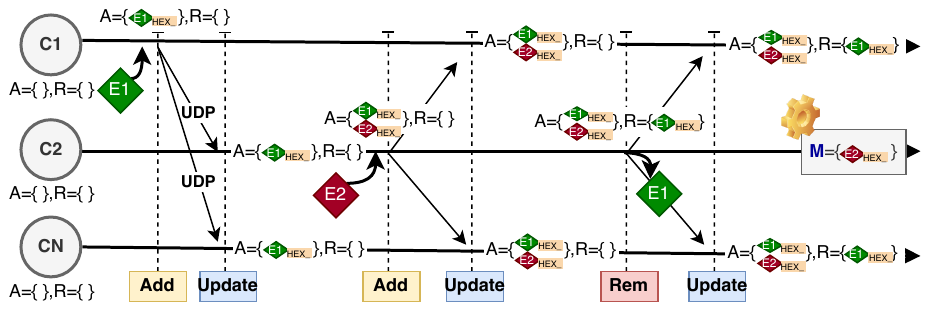}
    \caption{Conflict-free Replicated Set Data Operations.}
    \label{fig:setcvrddt}
\end{figure}

\begin{algorithm}
\caption{Update Set-CvRDT Protocol for Replicas.}
\label{alg:cvrdt_update}
\begin{algorithmic}[1]
\Require Received \texttt{add-set}($A_{rcv}$), \texttt{rem-set}($R_{rcv}$); Local $\texttt{add-set}$($A$), $\texttt{rem-set}$($R$), $\texttt{main-set}$($M$)

\ForAll{$item \in A_{recv}$}
    \If{$item \notin A$ \& $item \notin R_{recv}$}
        \State Insert $item$ into $M$ \Comment{Update main-set}
        \State Insert $item$ into $A$ \Comment{Track item addition}
    \EndIf
\EndFor

\ForAll{$item \in R_{recv}$}
    \If{$item \in \texttt{M}$}
        \State Remove $item$ from $M$ \Comment{Update main-set}
        \State Insert $item$ into $R$ \Comment{Track item removal}
    \EndIf
\EndFor
\end{algorithmic}
\end{algorithm}

\subsection{Distributed Conflict-free Replicated Set}
We implemented a conflict-free replicated data type (Set-CvRDT) for easy and reliable data sharing among distributed sensors. The set-CvRDT is essential in key areas of the \ourmethod{} framework. The data type is used to share SSL embeddings during clustering. Moreover, we store activity-affected sensor-cluster information to perform hyperlocal inference. The set-CvRDT has three grow-only sets: \textit{add-set} (A), \textit{rem-set} (R), and \textit{main-set} (M) as shown in the \figurename~\ref{fig:setcvrddt}. The figure illustrates consistent propagation of Set-CvRDT elements across multiple sensors (C1-CN) over an unreliable UDP communication channel. Each sensor begins with empty sets. When a sensor generates a set addition event (e.g., E1 at C1 or E2 at C2), it appends the (element E, unique HEX code) pair to its local add-set and broadcasts the update to others. The HEX code uniquely identifies each element in the replica sets. Due to UDP’s lossy and asynchronous nature, other sensors may receive the updates at different times. Each replica monotonically incorporates updates using the state update protocol in Algorithm~\ref{alg:cvrdt_update}, ensuring that all add operations eventually converge. When a removal occurs (e.g., removing E1), the originating sensor inserts the event into its rem-set (R) and disseminates the update. Finally, the main-set (M) at each sensor is updated from local and received copies of the add-set and rem-set, ensuring all replicas converge to the same set despite message delays, reordering, or loss.

The set-CvRDT implementation ensures crucial properties for distributed storage. We ensure all replicas will eventually have a consistent state (\textit{Convergence}). We ensure that failed or newly added nodes will ultimately reach the same final state as the other replicas (\textit{Fault Tolerance}). We ensure that the order or grouping of set operations (i.e., addition or removal of items) across replicas does not affect the final state (\textit{Commutativity} and \textit{Associativity}). 
Finally, we assign a unique HEX code to each item in the set to resolve conflicts, ensuring that each item can be reinserted (\textit{Reinsertions}) whenever required.

\subsection{Pollution-aware Clustering}
We implement pollution-aware sensor clustering using similarity-based partitioning and distance-aware merging within each partition. Each sensor computes its local SSL embeddings to represent its pollution context. When partitioning the network, we group clusters with their top nearest neighbors together. We compute the Euclidean distance of the SSL embeddings as a similarity measure between any two sensors. This forms the edges of the initial graph representation of the sensor network. We only include a list of edges with the shorter distance to threshold $\theta$, and represent all ignored edges as a lower bound $\texttt{b}_{L}$ indicating that their distances are greater than $\texttt{b}_{L}$. The distance-aware merging algorithm works on each partition and performs merges locally. Whenever it merges a cluster pair, it ensures that the two clusters are mutual nearest neighbors by checking distance bounds (i,e, $b_L$ and $b_U$), thereby guaranteeing the correctness of the clusters.

Given an undirected weighted graph \( G = (C, W) \), where \( C \) is a set of nodes and \( W \) is a set of weighted edges indicating the distances between pairs of items in \( C \), Algorithm~\ref{alg:cluster} initializes each node into its own cluster. We assume each initial node \( c \in C \) has an associated label denoted by \( \text{label}(c) \). Here, it's the sensor IDs. We further assume that the cluster label is the maximum label of the cluster's node. Let \( \textit{Dist}(C_i, C_j) \) be the euclidean distance between clusters \( C_i \) and \( C_j \), so that we can define weight \( w(C_i, C_j) \). We define \( L(C_i) \) as the list of nearest neighbors of \( C_i \), whose size limit is a configurable parameter. For each \( C_j \in L(C_i) \), we define \( b_L(C_i, C_j) \) and \( b_U(C_i, C_j) \) as the lower and upper bounds of \( \textit{dist}(C_i, C_j) \) respectively, both initialized to \( \textit{dist}(C_i, C_j) \). We update the distance between a merged cluster \( C_{ij} \) and any other cluster or node \( C_x \) as equation~\ref{equa:dist_link}. Similarly, distance bounds are also updated as per equation~\ref{equa:bl} and equation~\ref{equa:bu}.
\begin{align}
\label{equa:dist_link}
Dist(C_{ij}, C_x) &= \frac{Dist(C_i, C_x) \cdot |C_i| + Dist(C_j, C_x) \cdot |C_j|}{|C_i| + |C_j|}\\
\label{equa:bl}
b_L(C_{ij}, C_x) &= \frac{b_L(C_i, C_x) \cdot |C_i| + b_L(C_j, C_x) \cdot |C_j|}{|C_i| + |C_j|} \\
\label{equa:bu}
b_U(C_{ij}, C_x) &= \frac{b_U(C_i, C_x) \cdot |C_i| + b_U(C_j, C_x) \cdot |C_j|}{|C_i| + |C_j|}
\end{align}
\begin{figure}
        \centering
        \includegraphics[width=\columnwidth]{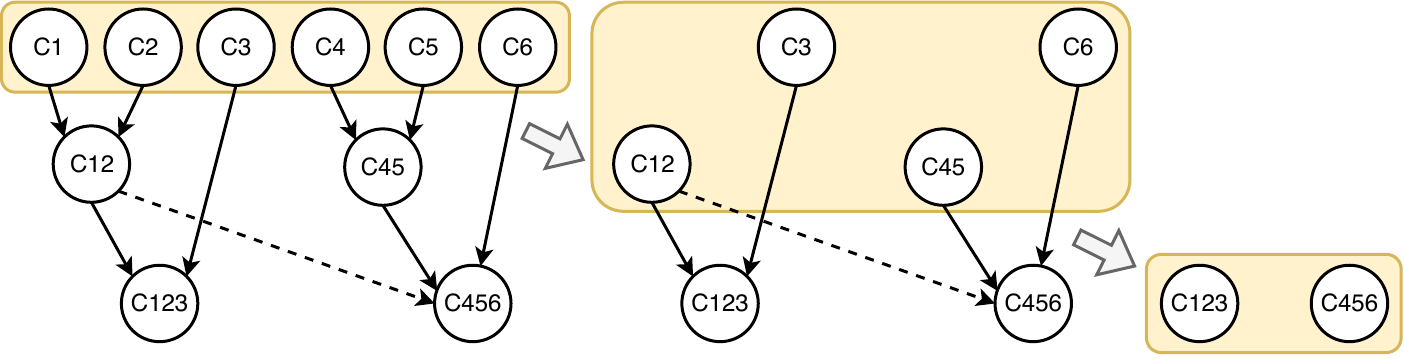}
        \caption{Pollution-aware merging of nodes in the clustering.}
        \label{fig:pack_example_fronteir}
\end{figure}

\begin{algorithm}
\caption{Pollution-aware Clustering of Sensor Nodes}
\label{alg:cluster}
\renewcommand{\algorithmicrequire}{\textbf{Input:}}
\renewcommand{\algorithmicensure}{\textbf{Output:}}
\begin{algorithmic}[1]

\Require Cluster graph $G = (C, W)$; Threshold $\theta$
\Ensure Final clusters $C^*$

\While{there exists $Dist(C_i, C_j) \leq \theta$}

    \State $P \gets Partition(G)$
    \State $C' \gets \emptyset$

    \ForAll{partitions $P_h = (\mathcal{C}_h, \{\mathcal{L}(C_i)\mid C_i \in \mathcal{C}_h\}) \in P$}

        \State $/*$ Polluiton-aware local merging for $P_h$ $*/$

        \State $G_h \Leftarrow$ Build a graph from $P_h$

        \ForAll{$C_i \in \mathcal{C}_h$}
            \State $NN(C_i) \Leftarrow$ nearest neighbor of $C_i$ whose upper bound ($b_U$) 
            is smaller than the lower bounds ($b_L$) of other neighbors; $null$ if non-existent
        \EndFor

        \While{$true$}
            \If{$\exists(C_i, C_j) : (NN(C_i)=C_j) \wedge (NN(C_j)=C_i) \wedge (b_{U}(C_i,C_j)\le \theta)$}
                \State Merge $C_i$ with $C_j$ in $G_h$, 
                \State Update $G_h$, and $NN(\cdot)$
            \Else
                \State \textbf{break}
            \EndIf
        \EndWhile

        \State $C' \gets C' \cup$ Merged clusters in $G_h$
    \EndFor

    \State $C \gets$ Integrate $C'$ \Comment{Merge overlapping clusters}
    \State $W \gets$ Merge weights based on $C$

\EndWhile

\State $C^* \gets C$

\end{algorithmic}
\end{algorithm}

We merge clusters within a partition as long as there exists a pair of mutual nearest neighbors \( (C_i, C_j) \), and \( b_U(C_i, C_j) \leq \theta \) as per lines 11 to 13 in the Algorithm~\ref{alg:cluster} (See \figurename~\ref{fig:pack_example_fronteir}). It outputs a set of clusters \( C' \). Lastly, distance bounds are used to merge overlapping clusters before the final clusters \( C^* \) output.

\subsection{Hyperlocal Inference}
Each activity-affected sensor cluster independently performs hyperlocal inference to determine the specific indoor activity occurring within its localized region. To coordinate this process without relying on a central controller, each cluster initiates a second RAFT-based leader-election round~\cite{ongaro2014raft}, ensuring that a single inference leader is selected among the affected sensors. Once elected, the inference leader collects SSL embeddings from all cluster members that already encode the time–frequency characteristics of local air-quality fluctuations, and aggregates them for classification. The leader then executes inference using off-the-shelf ML models (i.e., Decision Trees, Random Forests, Extra Trees, Gaussian Naive Bayes, and lightweight Neural Networks). By assigning inference responsibilities only to the sensors affected by the localized activity, the system achieves fine-grained, energy-efficient HAR while preserving the spatial relevance of the indoor pollutants.

\section{Experimental results}
Here, we present experimental results with \ourmethod{} framework over our in-house pollution sensor network deployment.

\subsection{Implementation Details}
\noindent\textbf{Sensor Network Deployment:}
Our study deployed a dense network of low-cost air quality sensors across $30$ diverse indoor sites over six-months, covering both summer and winter seasons. These sites include studio apartments, shared classrooms, research laboratories, residential households, and food canteens, providing a representative ecosystem of real-world human activity and pollution dynamics. Across the deployment, the sensing infrastructure collected $89.1$ million samples, amounting to $13646$ hours of continuous indoor air quality measurements. To contextualize the sensor data with human behavior, we incorporated $3957$ manually annotated activity events from $24$ active participants (among $46$ total occupants). These annotations capture a wide spectrum of indoor activities, including engagement and occupancy (i.e., enter, exit), occupant behavior (i.e., fan on/off, AC on/off), prohibited or impactful practices (i.e., gathering, eating), and cooking-related activities: annotated across five cooking categories (i.e., boiling, deep-frying, shallow-frying, steaming, and shallow-frying with boiling) and eleven food items (e.g., tea, rice, fish, egg, lentils, leafy vegetables). This large-scale, heterogeneous deployment enables a fine-grained analysis of spatiotemporal indoor pollution patterns shaped by everyday human activities and diverse cooking practices.

\noindent\textbf{SSL Model and Off-the-shelf ML Models:}
To implement the Self-supervised embedding network, we adapted the neural network architecture from the time-frequency consistency model~\cite {zhang2022self}. Further, we have implemented Decision Tree (DT), Random Forest (RF), Extra Trees (ET), Gaussian Naive Bayes (NB), and Multilayer Perceptron (MLP) classifiers to detect hyperlocal activities at the leader nodes with the \textit{emlearn}\cite{emlearn} Python library.

\noindent\textbf{Evaluation Metric:}
We evaluate the platform using four key metrics. Latency is measured in microseconds, milliseconds, or seconds to capture responsiveness across system operations. Memory usage is tracked in bytes to assess the platform’s footprint on resource-constrained ESP32 devices. Power consumption is measured in milliwatts to quantify energy efficiency during different stages of \ourmethod{}. Finally, model performance is evaluated based on accuracy in predicting indoor activities. These metrics collectively provide a clear view of system efficiency and predictive reliability.

\subsection{SSL Embeddings Quality Analysis}
\begin{figure}  
    \centering  
    \includegraphics[width=1.0\columnwidth]{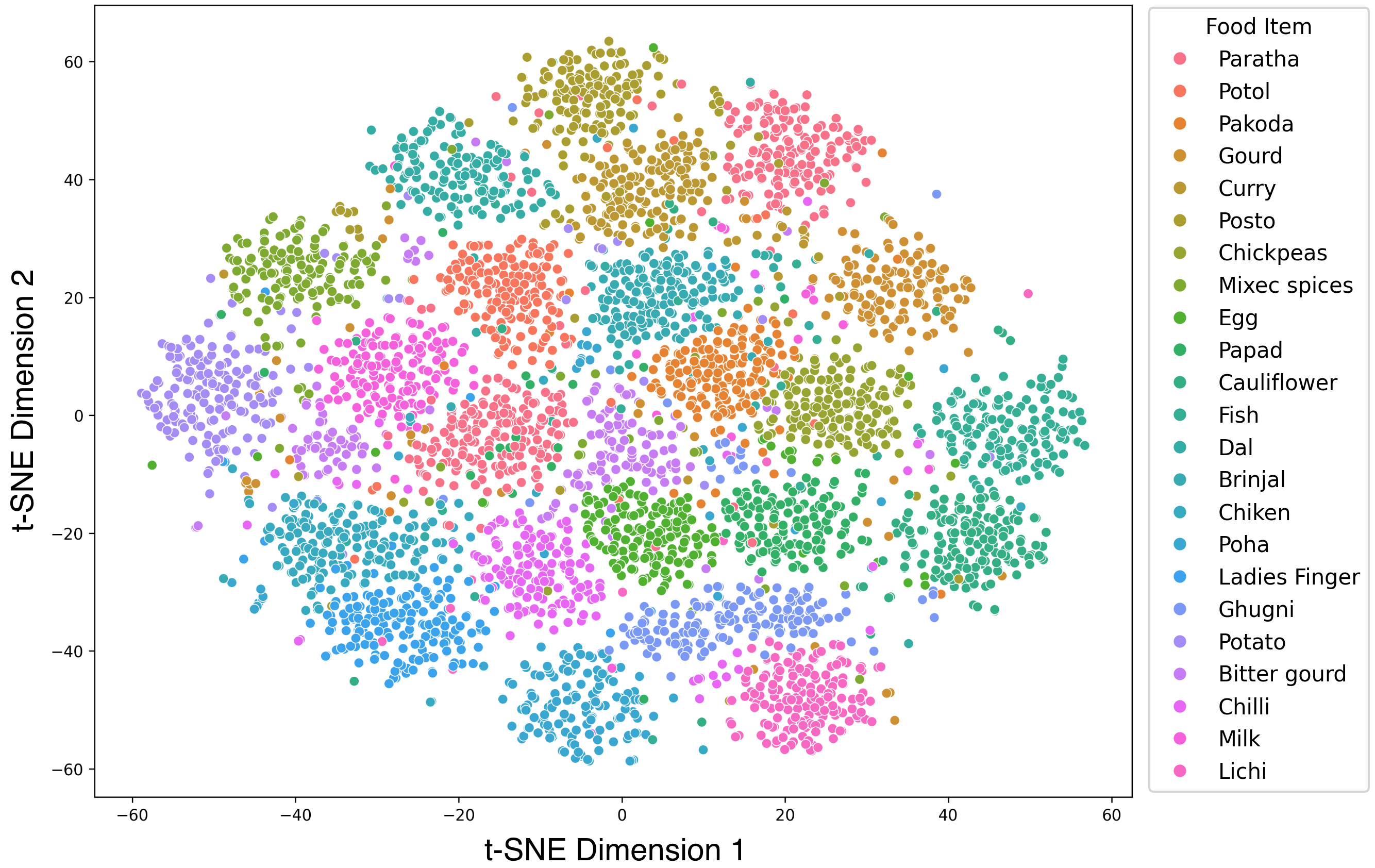}  
    \caption{t-SNE visualization of cooked food items.}  
    \label{fig:tfc_tsne}  
\end{figure}

Figure \ref{fig:tfc_tsne} illustrates the SSL-based latent embeddings of the cooking activities projected onto a two-dimensional space by applying t-SNE to the embeddings extracted from the trained time-frequency consistency (TF-C) model. The plot shows that the TF-C model successfully captured meaningful semantic relationships between different food items, resulting in 24 distinct clusters. Each food item category, including Paratha, Potol, Pakoda, Gourd, Curry, Posto, Chickpeas, Mixed spices, Egg, Papad, Cauliflower, Fish, Dal, Brinjal, Chicken, etc., forms a relatively cohesive cluster in the embedding space.

\subsection{Pollution-aware Clustering Effeciency}

\begin{figure*}
	\captionsetup[subfigure]{}
	\begin{center}
    \subfloat[Memory with nodes\label{fig:pack_mem}]{
			\includegraphics[width=0.24\linewidth,keepaspectratio]{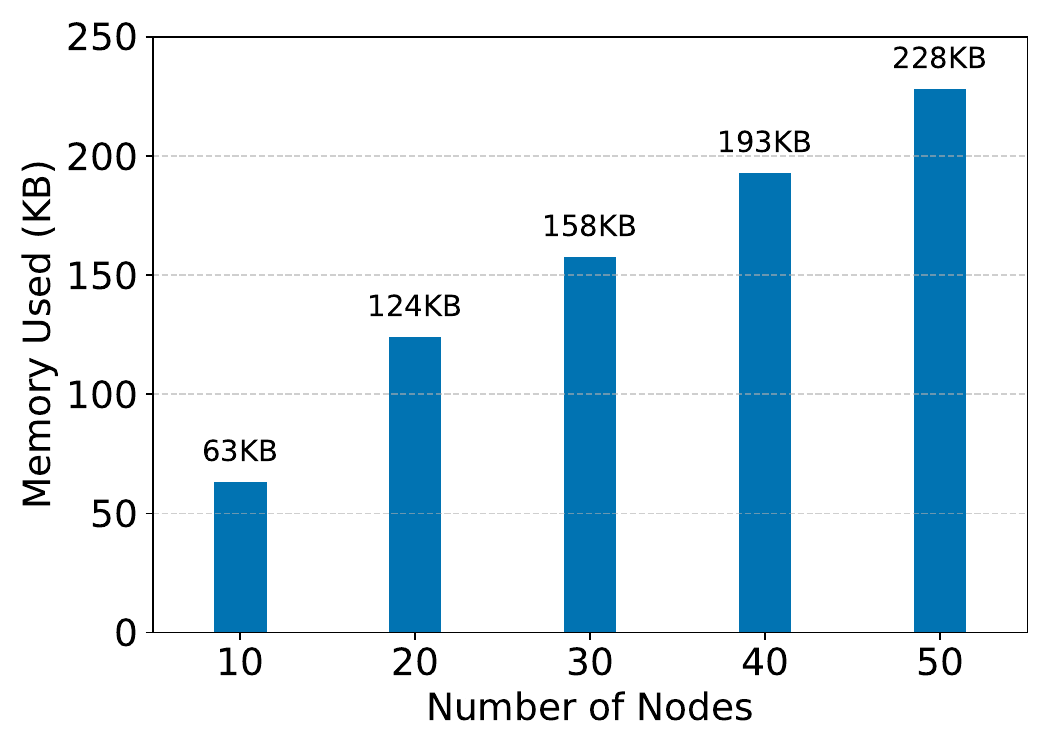}
		}
            \subfloat[Power usage with nodes\label{fig:pack_power_box}]{
			\includegraphics[width=0.24\linewidth,keepaspectratio]{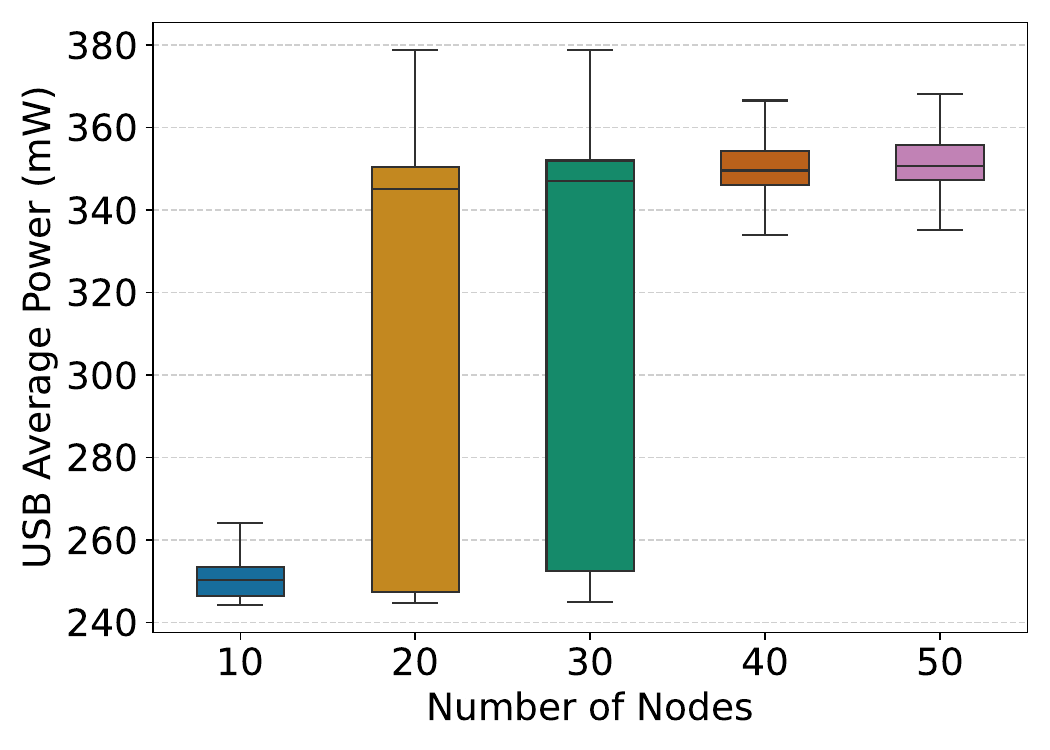}
		}
            \subfloat[Power draw\label{fig:pack_power_line}]{
			\includegraphics[width=0.24\linewidth,keepaspectratio]{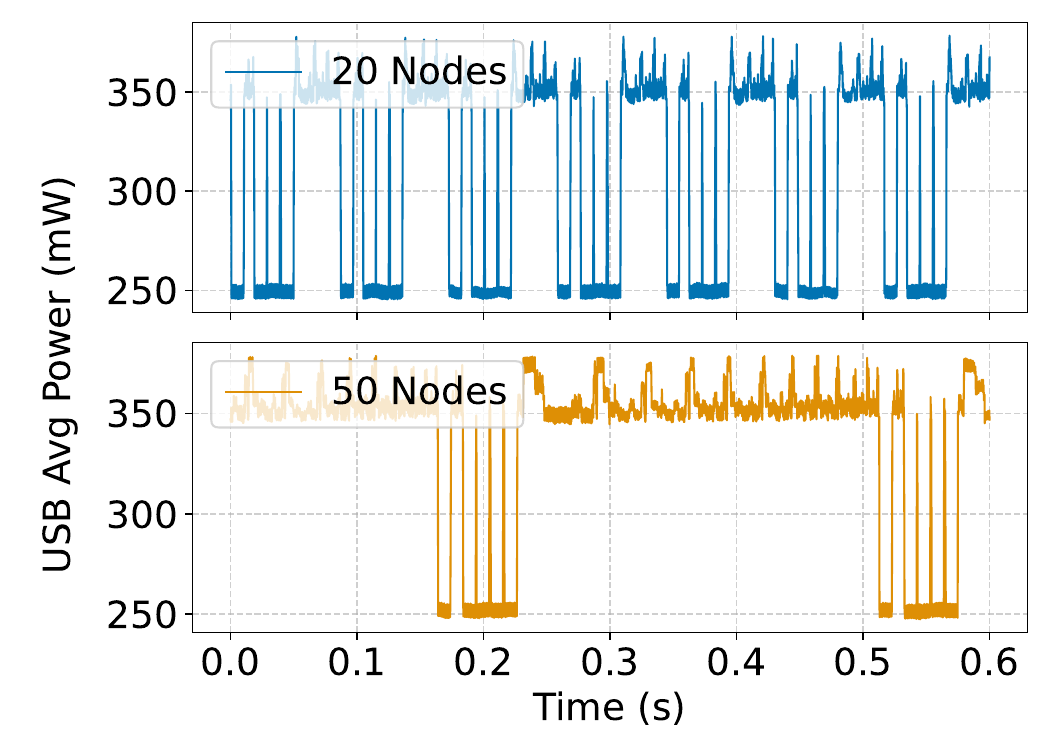}
		}
        \subfloat[Clustering time with nodes\label{fig:pack_exe_time}]{
			\includegraphics[width=0.24\linewidth,keepaspectratio]{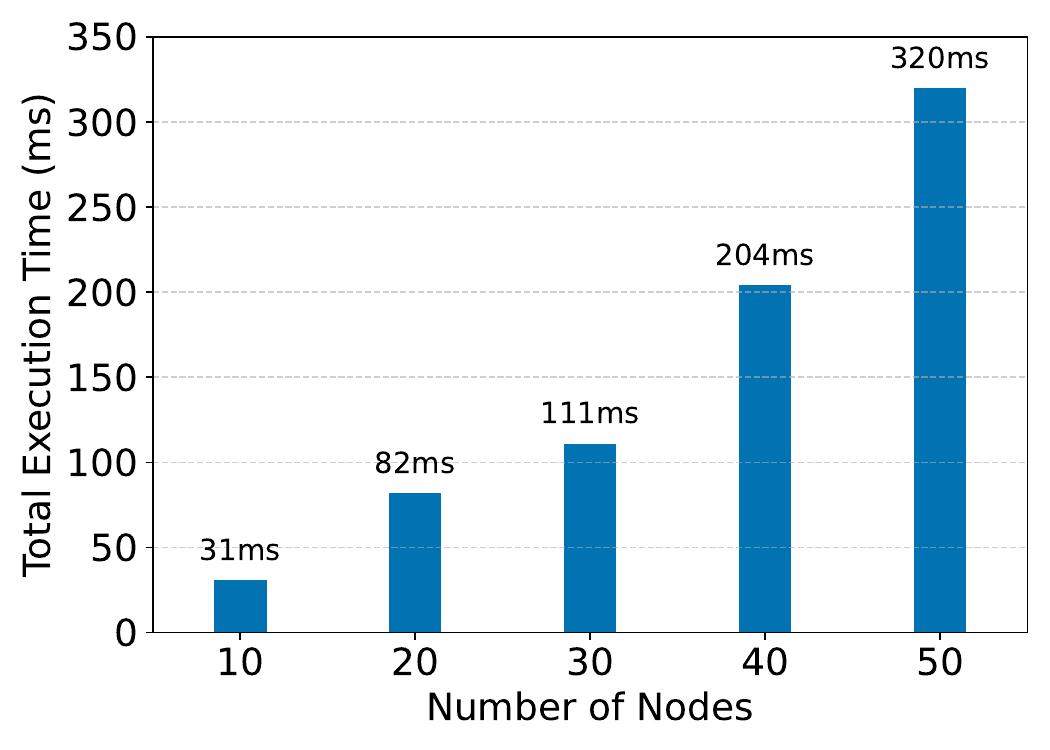}
		}
	\end{center}
	\caption{Pollution-aware clustering performance: (a) memory usage with number of nodes, (b) power usage with number of nodes, (c) power draw while execution, (d) total clustering time with number of nodes.}
	\label{fig:pack_power}
\end{figure*}

\noindent\textbf{Memory Usage:} 
Memory consumption increases nearly linearly with the number of nodes, rising from approximately $63$ KB at $10$ nodes to about $228$ KB at $50$ nodes, as shown in \figurename~\ref{fig:pack_mem}. This reflects the overhead of maintaining additional cluster structures, distance bounds, and nearest-neighbor mappings as the system scales. The steady growth demonstrates a lightweight memory footprint suitable for \ourmethod{}.

\noindent\textbf{Power Usage with Number of Nodes:} 
Average power usage shows a gradual upward trend with increasing nodes, with median USB power consumption ranging from roughly $250$ mW ($10$ nodes) to $360$ mW ($50$ nodes) as per \figurename~\ref{fig:pack_power_box}. Despite the increase, the boxplots indicate stable, bounded power behavior, suggesting that clustering computations introduce minimal energy variability in low-power deployments.

\noindent\textbf{Real-Time Power Draw During Execution:} 
Power traces in \figurename~\ref{fig:pack_power_line} exhibit periodic bursts correlating with cluster-merge and graph-update operations. For both cases, power fluctuates between $260$--$360$ mW, while for $50$ nodes, peaks become slightly less frequent due to increased clustering time at the leader. These patterns confirm that the algorithm performs short, compute-intensive bursts rather than sustained heavy loads, making it compatible with real-time execution.

\noindent\textbf{Clustering Time with Increasing Nodes.} 
As observed in the power analysis, execution time scales proportionally with the number of nodes, increasing from $31$ ms at $10$ nodes to $320$ ms at $50$ nodes as shown in \figurename~\ref{fig:pack_exe_time}. Even at the largest configuration, the total clustering time remains well below one second, highlighting the efficiency of partition-based processing and incremental cluster-graph updates. This characteristic is crucial for pollution-aware systems using multiple distributed sensors.

\begin{figure}
	\captionsetup[subfigure]{}
	\begin{center}
            \subfloat[RAFT power usage\label{fig:raft_power_boxplot}]{
			\includegraphics[width=0.48\linewidth,keepaspectratio]{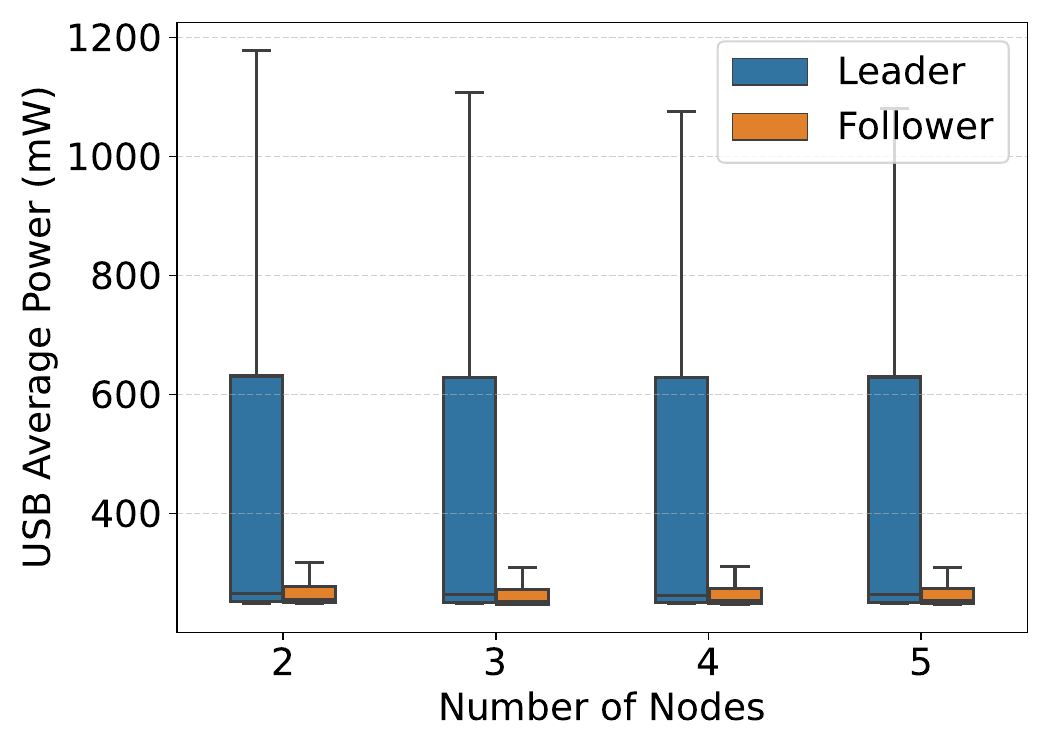}
		}
            \subfloat[Recovery time on leader failure\label{fig:raft_recovery_time}]{
			\includegraphics[width=0.48\linewidth,keepaspectratio]{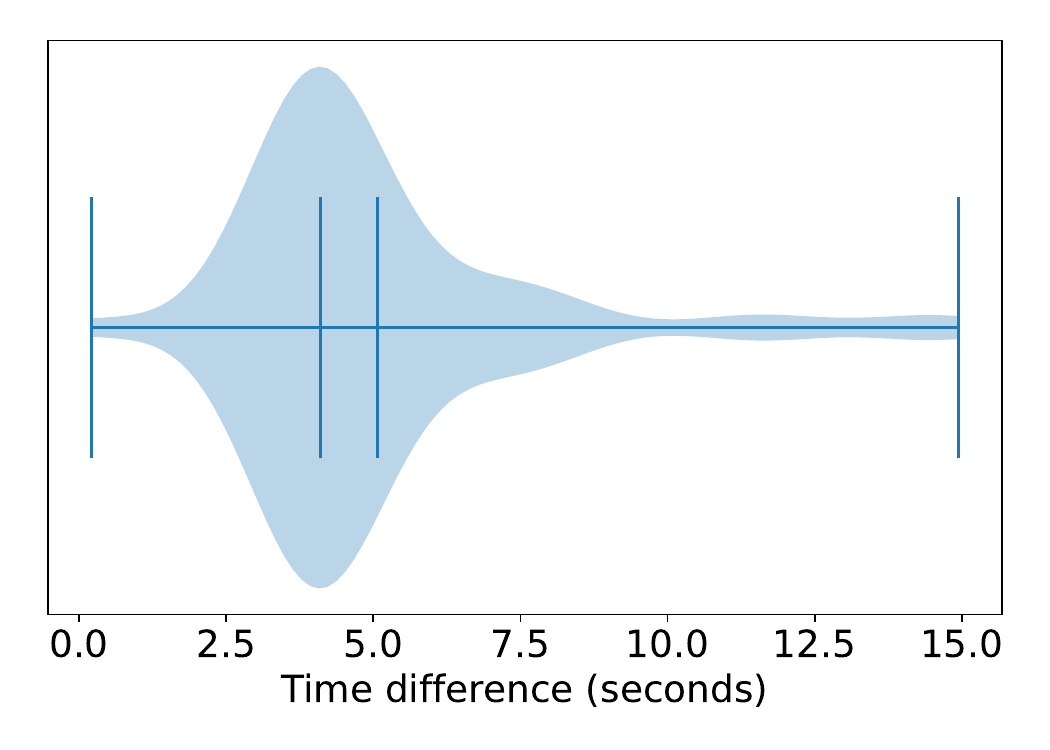}
		}
	\end{center}
	\caption{RAFT-based leader election performance: (a) power usage with number of cluster nodes, (b) Recovery time to find a new cluster leader on current leader failure.}
	\label{fig:RAFT_statistics}
\end{figure}

\subsection{Distributed Leader Election Analysis}
\noindent\textbf{Leader and Follower Power Consumption:}
\figurename~\ref{fig:raft_power_boxplot} compares the power usage of leader and follower nodes across cluster sizes ranging from two to five devices. The leader consistently consumes more power (approximately $600$--$650$ mW) due to its coordination responsibilities, including heartbeat generation and log management. In contrast, follower nodes operate at a lower and stable power range ($270$--$300$ mW), with negligible variation as the number of nodes increases. This indicates that RAFT imposes minimal incremental overhead as clusters grow, making it efficient for resource-constrained settings.

\noindent\textbf{Leader Failure Recovery Time:}
\figurename~\ref{fig:raft_recovery_time} presents the temporal distribution of leader recovery times following induced leader failures. The RAFT protocol reliably re-establishes the leader within approximately $3$--$6$ seconds. The narrow spread of the violin plot reflects consistent timeout-triggered re-election across repeated trials, demonstrating RAFT’s robustness and predictable fault tolerance in the sensor network.

\begin{figure*}
	\captionsetup[subfigure]{}
	\begin{center}
            \subfloat[Indoor activity\label{fig:CM_RF_lab}]{
			\includegraphics[width=0.32\linewidth,keepaspectratio]{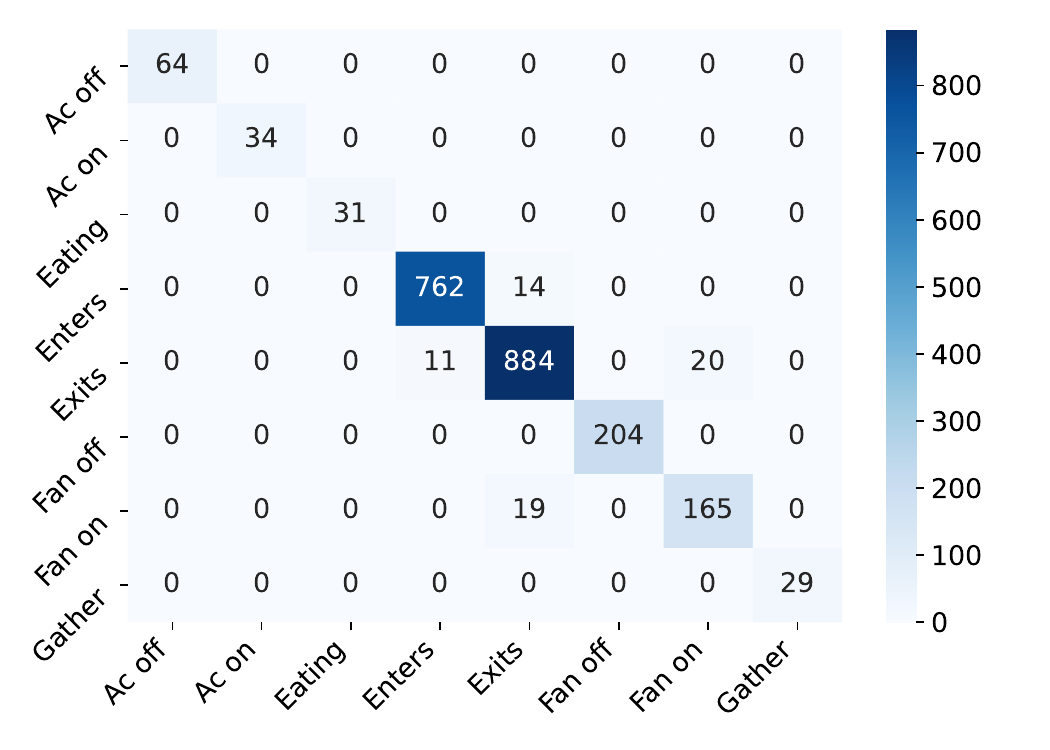}
		}
            \subfloat[Cooking type\label{fig:CM_RF_CTYPE}]{
			\includegraphics[width=0.32\linewidth,keepaspectratio]{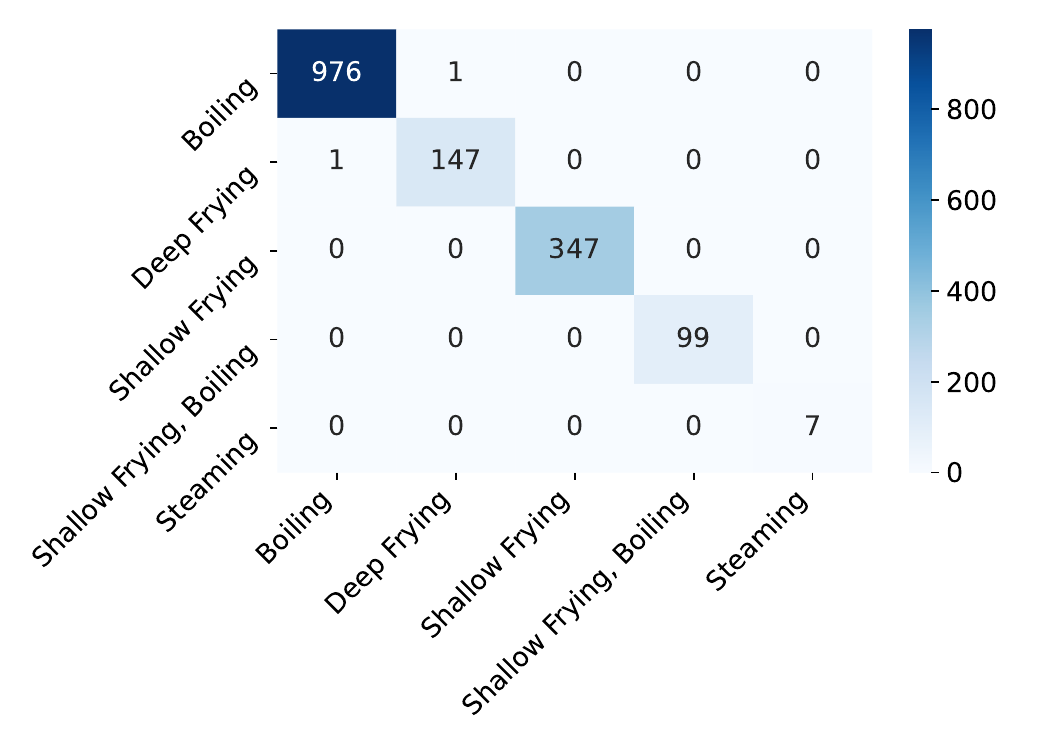}
		}
            \subfloat[Food Item\label{fig:CM_RF_FTYPE}]{
			\includegraphics[width=0.32\linewidth,keepaspectratio]{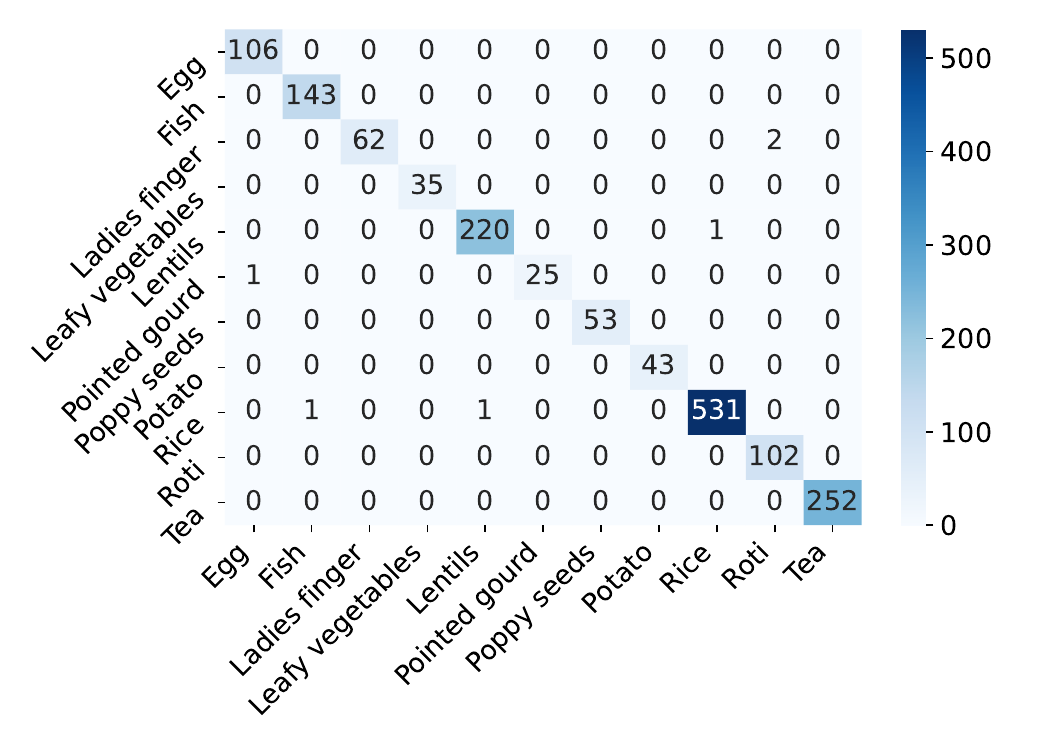}
		}
	\end{center}
	\caption{Classification confusion matrix of Random Forest: (a) indoor activity dataset, (b) cooking type, and (c) food item in the pollution sensor network deployment.}
	\label{fig:RF_perf}
\end{figure*}

\subsection{Activity Classification Performance}
\noindent\textbf{Detection of Indoor Activities}
Tree-based models exhibited the strongest performance across all metrics, with \textit{Random Forest emerging as the best overall model}, achieving an accuracy of 97.41\% with an average inference time of only 34\,\textmu s as shown in \tablename~\ref{tab:model_performance}. The confusion matrix is shown in \figurename~\ref{fig:CM_RF_lab}. Extra Trees offered comparable accuracy but incurred higher latency, while the Decision Tree remained the fastest (9\,\textmu s) with competitive accuracy. In contrast, models relying on floating-point operations, such as Gaussian Naive Bayes and the small Neural Network, performed poorly on the ESP32, yielding less than 50\% accuracy and significantly longer prediction times, due to limited floating-point compute on ESP32.

\begin{table}[h]
\centering
\scriptsize
\caption{Model Performance for Indoor Activities.}
\begin{tabular}{|l|c|c|c|c|}
\hline
\textbf{Model} & \textbf{Accuracy} & \textbf{Avg. Time} & \textbf{Max Time}  \\
\hline
Decision Tree & 93.56\% & 9 µs & 42 µs \\
Random Forest & 97.41\% & 34 µs & 330 µs \\
Extra Trees & 97.36\% & 62 µs & 463 µs \\
Gaussian NB & 48.28\% & 356 µs & 382 µs \\
Neural Network & 49.45\% & 89 µs & 438 µs \\
\hline
\end{tabular}
\label{tab:model_performance}
\end{table}

\noindent\textbf{Detection of Cooking Activities}
Across both cooking type and food identification tasks, tree-based models consistently delivered the strongest results. \textit{Random Forest emerged as the most effective model overall}, achieving accuracy above 99.4\% with low inference times of 16--23\,\textmu s, offering the best balance between accuracy and computational cost as shown in \figurename~\ref{fig:RF_perf}, \tablename~\ref{tab:cooking_type_performance}, and \tablename~\ref{tab:food_type_performance}. Extra Trees achieved slightly higher accuracy (up to 99.81\%) but required considerably more time (43--64\,\textmu s), while Decision Tree provided the fastest predictions (6--7\,\textmu s) with strong accuracy for both tasks. All tree-based models maintained real-time inference capability on ESP32.

In contrast, models that relied heavily on floating-point operations, Neural Networks, and Gaussian Naive Bayes performed poorly on the ESP32 platform. Even with a reduced neural network architecture (two 10-neuron hidden layers), accuracy dropped below 50\% and inference times increased substantially, largely due to ESP32’s limited floating-point support and the required integer conversion of sensor data. Overall, \textbf{Random Forest proved to be the optimal choice}, delivering high accuracy with efficient computation, enabling real-time cooking activity detection.

\begin{table}[h]
\centering
\scriptsize
\caption{Model Performance for Cooking Classification.}
\begin{tabular}{|l|c|c|c|c|}
\hline
\textbf{Model} & \textbf{Accuracy} & \textbf{Avg. Time} & \textbf{Max Time}  \\
\hline
Decision Tree & 97.28\% & 6 µs & 39 µs \\
Random Forest & 99.68\% & 16 µs & 271 µs \\
Extra Trees & 99.81\% & 43 µs & 339 µs \\
Gaussian NB & 47.72\% & 157 µs & 181 µs \\
Neural Network & 44.36\% & 61 µs & 352 µs \\
\hline
\end{tabular}
\label{tab:cooking_type_performance}
\end{table}

\begin{table}[h]
\centering
\scriptsize
\caption{Model Performance for Food Item Classification.}
\begin{tabular}{|l|c|c|c|c|}
\hline
\textbf{Model} & \textbf{Accuracy} & \textbf{Avg. Time} & \textbf{Max Time}  \\
\hline
Decision Tree & 94.17\% & 7 µs & 38 µs \\
Random Forest & 99.49\% & 23 µs & 289 µs \\
Extra Trees & 99.68\% & 64 µs & 311 µs \\
Gaussian NB & 46.52\% & 178 µs & 209 µs \\
Neural Network & 45.71\% & 83 µs & 391 µs \\
\hline
\end{tabular}
\label{tab:food_type_performance}
\end{table}

\begin{figure}
	\captionsetup[subfigure]{}
	\begin{center}
            \subfloat[ML model average power usage\label{fig:ml_model_power}]{
			\includegraphics[width=0.48\linewidth,keepaspectratio]{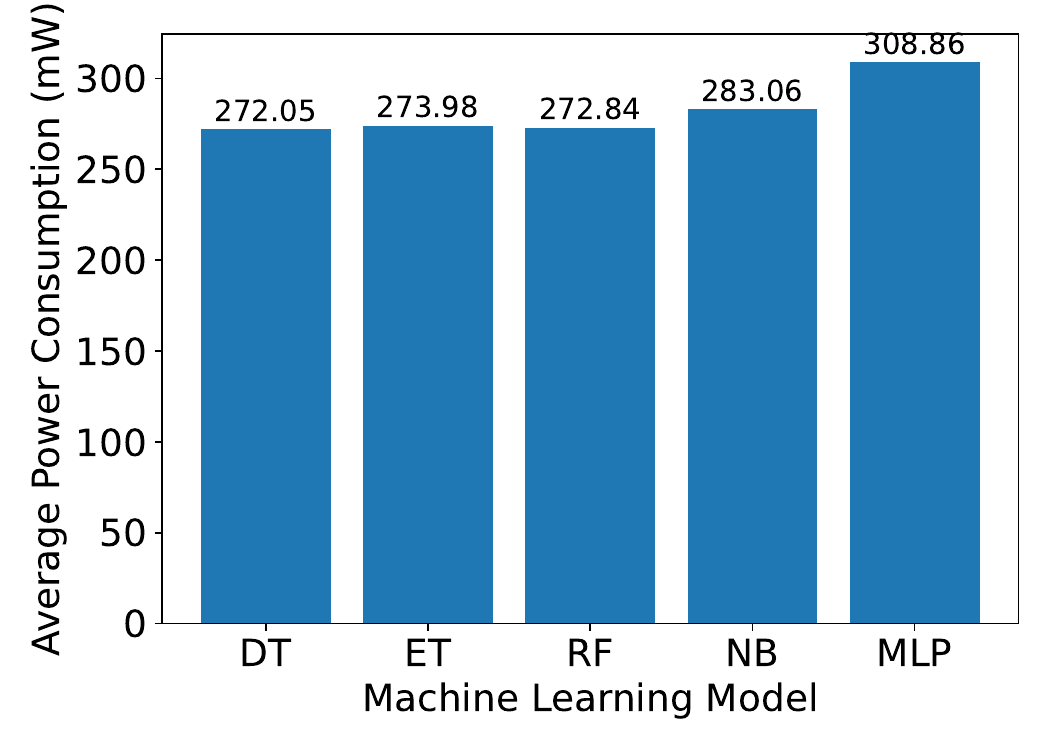}
		}
            \subfloat[ML model Power draw\label{fig:ml_power_line}]{
			\includegraphics[width=0.48\linewidth,keepaspectratio]{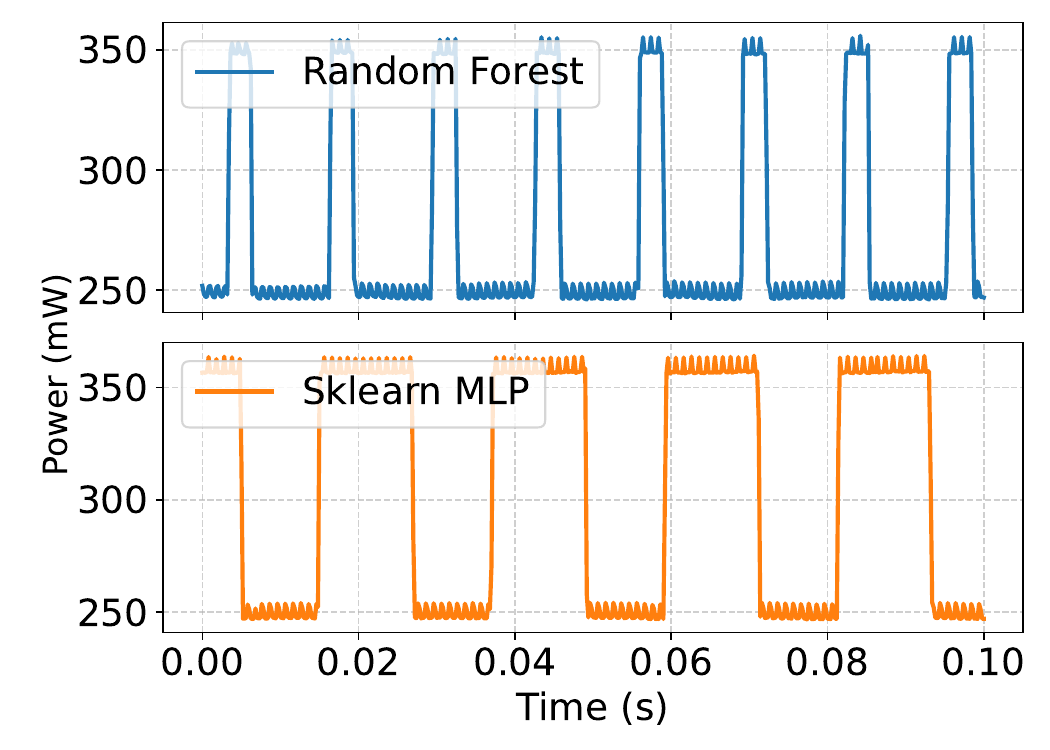}
		}
	\end{center}
	\caption{On-device power usage for hyperlocal inference.}
	\label{fig:ML_statistics}
\end{figure}

\subsection{On-device Inference Power Analysis}

\figurename~\ref{fig:ml_model_power} compares the average power consumption of different ML models during on-device inference. Tree-based models (i.e., Decision Tree, Extra Trees, and Random Forest) show nearly identical power usage of $272$--$274$ mW, indicating that their lightweight integer-based computations impose minimal additional load on the ESP32. Gaussian Naive Bayes incurs slightly higher power consumption ($283$ mW), likely due to repeated floating-point operations. In contrast, the MLP model consumes the highest power at $308$ mW, reflecting the computational overhead of neural network inference on constrained hardware.

\figurename~\ref{fig:ml_power_line} presents fine-grained power draw traces for Random Forest and the MLP. Both models exhibit periodic spikes corresponding to inference cycles, but their magnitudes and durations differ significantly. Random Forest exhibits short, sharp spikes with a quick return to idle, illustrating its fast execution and low computational burden. The MLP, however, shows longer, higher-power spikes, indicating prolonged processing times and heavier compute demands. These patterns reinforce that tree-based models not only deliver better accuracy but also maintain more energy-efficient real-time performance for hyperlocal inference on ESP32.

\section{Conclusion}
We presented \ourmethod{}, a lightweight and scalable framework that enables distributed air-quality sensor networks to collaboratively detect hyperlocal indoor activities under strict hardware constraints. By integrating CRDT-based data sharing, hierarchical clustering with a self-supervised distance metric, and RAFT-style leader-based group inference, our approach effectively addresses the core challenges of on-device processing, consensus maintenance, and activity-aware sensor grouping in low-power deployments. Extensive evaluation on real-world datasets demonstrates that \textsc{PoHAR} achieves high inference accuracy of 97.41\% for indoor activities and 99.68\% for cooking activities, while maintaining below 34 \textmu s latency and milliwatt-level power usage. The results highlight the viability of repurposing commodity air-quality sensors for privacy-preserving human activity recognition. Open-sourced at \url{https://github.com/prasenjit52282/PoHAR}.
\section*{acknowledgement}
The work is supported by the Prime Minister Research Fellowship (IIT/Acad/PMRF/SPRING/2022-23, dated 24 March 2023) and Google Award on Society-Centered AI 2025.

\bibliographystyle{ieeetr}
\bibliography{reference}

\balance

\end{document}